\documentclass[aps,prl,twocolumn,a4paper,10pt]{revtex4-1}

\usepackage[T1]{fontenc}		
\usepackage[english]{babel}		
\usepackage[latin1]{inputenc}	
\usepackage{times}

\usepackage{amsmath}			
\usepackage{amssymb}			
\usepackage{bbm}				
\usepackage{mathtools}
\usepackage{simpler-wick}
\usetikzlibrary{tikzmark}
\DeclarePairedDelimiterX\Dirbraket[3]{\langle}{\rangle}%
{#1\,\delimsize\vert\,\mathopen{}#2\,\delimsize\vert\,\mathopen{}#3}

\usepackage[colorlinks=true, a4paper=true, pdfstartview=FitV,linkcolor=blue, citecolor=blue, urlcolor=blue]{hyperref}

\usepackage{graphicx}			
\usepackage{graphics}			
\DeclareGraphicsExtensions{.pdf}
\usepackage{color}		

\newcommand{\bea}{\begin{eqnarray}}
\newcommand{\eea}{\end{eqnarray}}

\newcommand{\bk}{\mathbf{k}}

\newcommand{\Tr}{\text{Tr}}

\begin{document}

\title{Strong-coupling diagrammatic Monte Carlo technique for correlated fermions and frustrated spins}

\author{Johan~Carlstr\"om }
\affiliation{Department of Physics, Stockholm University, 106 91 Stockholm, Sweden}
\date{\today}

\begin{abstract}
We describe a controllable and unbiased strong-coupling diagrammatic Monte Carlo technique that is applicable to a wide range of fermionic systems and spin models. 
Unlike previous strong coupling methods that generally rely on the Grassmannian Hubbard-Stratonovich transformation, our construction is based on Wick's theorem and a recursive procedure to group contractions into effective connected vertices that are non-perturbative in all local physics and can be calculated exactly. 
The resulting expansion is described by simple diagrammatic rules that make it suitable for systematic treatment via stochastic sampling. Benchmarks against numerical linked cluster expansion display excellent agreement. 
\end{abstract}
\maketitle


Strongly correlated electrons and frustrated spin models are among the most challenging problems in condensed matter theory due to a combination of the sign problem, lack of a natural small parameter and the computational complexity of series expansions. 
 Coincidentally, this topic is at the same time absolutely central to understanding the electronic structure of solids, and a wide range of numerical techniques have accordingly been devised to overcome these obstacles.
 A well-known example of this is DMFT \cite{PhysRevLett.62.324}, with extensions based on diagram techniques, \cite{PhysRevB.75.045118,PhysRevB.77.033101}, cluster generalizations \cite{PhysRevLett.106.047004,PhysRevLett.110.216405,RevModPhys.77.1027}, and related methods \cite{PhysRevLett.112.196402,PhysRevB.92.115109}.
 Other examples include DMRG \cite{PhysRevLett.69.2863}, wave function methods 
 \cite{PhysRevLett.87.217002,PhysRevB.95.024506}, and auxiliary-field quantum Monte Carlo \cite{PhysRevB.39.839,PhysRevLett.62.591,PhysRevB.40.506,afqmc}.

Several of the aforementioned techniques are capable of producing states that seem highly relevant for cuprate superconductivity, including anti-ferromagnetism, stripes, pseudogap physics, and d-wave superconductivity. Nevertheless, it has been known for some time that notable discrepancies may appear both when comparing different techniques and when altering details of the implementation (such as discretization) of a given method \cite{2006cond.mat.10710S}.
 This sensitivity that correlated-fermion models display to numerical protocol appears to have a physical origin, and be rooted in competition between different states situated very closely in terms of free energy \cite{2006cond.mat.10710S,Dagotto257}. 
 
 More recently, systematic comparison of leading numerical protocols, applied to the Hubbard model, has demonstrated some of the very significant progress that has eventually been made in this field \cite{PhysRevX.5.041041}. In a substantial part of the parameter space, key observables can now be obtained from completely different techniques with a high degree of consensus. The region that remains the most problematic corresponds to small but finite doping, and intermediate to large onsite repulsion. Incidentally, this is also a scenario that is highly relevant for cuprate superconductors.

The sensitivity that correlated systems display to perturbations suggests that a reliable solution to this problem requires numerical methods that can provide extremely accurate results in the strongly correlated regime, and this has proven to be a significant challenge.
The unbiased methods--that are free of systematic errors beyond some form of truncation--are typically based on a series expansion of some form. However, for strongly correlated systems, conventional perturbation theory is not viable, as the interaction is far too strong. Instead, an alternative expansion parameter has to be found.

The perhaps most well-known formalism aimed at the correlated regime is strong-coupling expansion, where the nonlocal processes are treated as a perturbation while the unperturbed system corresponds to the atomic limit \cite{PhysRevD.19.1865,PhysRevLett.80.5389}. Thus far however, most of these works include only modest expansion orders, and they are primarily conducted at half-filling, or for actual spin systems \cite{2011ifm..book..537M}, while the case of non-zero doping is technically far more challenging \cite{Pairault2000}. 

  Numerical linked cluster expansion (NLCE) \cite{PhysRevLett.97.187202} has been applied successfully to both spin models \cite{PhysRevE.75.061118} and itinerant fermionic theories like the t-J \cite{PhysRevE.75.061119} and Hubbard models.
  For the latter, results exist at infinite onsite repulsion \cite{PhysRevE.89.063301}, and for finite interactions up to $U/t=16$ \cite{PhysRevA.84.053611}, which is far into the strongly correlated regime. 
  This method is based on exact diagonalization of small clusters, and allows convergence to macroscopic results to be observed with increasing cluster size. 

 More recently, the extremely correlated fermi liquid theory was developed specifically for Gutzwiller-projected models \cite{PhysRevB.81.045121}. This framework allows a form of diagram technique to be employed on restricted Hilbert spaces \cite{PEREPELITSKY20151}, and currently published benchmarks, while limited to low expansion orders, appear encouraging \cite{PhysRevB.98.205106}.

Finally, a number of adaptions of diagrammatic Monte Carlo methods \cite{Van_Houcke_2010} have been made to address the strongly correlated regime. 
 Universal fermionization \cite{PhysRevB.84.073102} has opened a new analytical path where restrictions on the Hilbert space are encoded via non-Hermitian terms in the Hamiltonian, thus allowing  Gutzwiller-projected systems to be treated within the framework of Wick's theorem. 
Via second fermionization, doubly occupied sites can then be reintroduced in the form of hardcore bosons, which are subsequently fermionized, thus allowing generic correlated systems to be addressed within this framework \cite{PhysRevB.84.073102}. 
This technique suffered from poor convergence properties at its inception except at large doping, but this problem has since been overcome through spin-charge transformation, which essentially results in a representation involving fermionic carriers that propagate on a spin background \cite{0953-8984-29-38-385602}.
Diagrammatically, these models can be treated as a spin system using Popov-Fedotov fermionization \cite{JETP.67.535}, where the spins are mapped onto fermions with an imaginary chemical potential. Results from spin-charge transformation diagrammatic Monte Carlo (SCT-DMC) indicate that the expansion converges quite rapidly, but that the complexity of the resulting theory limits the expansion order, making it hard to reach low temperatures \cite{PhysRevB.97.075119}.

Currently, several new analytical techniques are being evaluated to overcome the inherent problem of a large expansion parameter: 
By extracting the analytical structure of the self-energy from weak-coupling data, it becomes possible to reconstruct it in the non-perturbative regime \cite{PhysRevB.100.121102}. Homotopic action operates on the principle of altering the starting point of the expansion, as well as the expansion parameter, such that the point of interest falls within the convergence radius \cite{kim2020homotopic}.
Taking advantage of determinant sampling techniques, \cite{PhysRevLett.119.045701,Rossi_2017}, these methods give access to fairly strong interactions, up to $U/t= 7$ in the Hubbard model. While impressive, this is still short of $U/t\sim 12$, which is relevant for the high-temperature superconductors.


Thus, despite the significant recent advances, much of the parameter space remains challenging to unbiased techniques, and it is the case of strong correlation that remains the most elusive. 
In this work, we will discuss how diagrammatic techniques can be adapted to the strongly correlated regime by an alternative series expansion that is based on a non-perturbative treatment of all the interactions, and expansion only in the nonlocal part of the Hamiltonian. 
This series expansion is computationally economical, possesses simple diagrammatic rules, and is free of any large expansion parameter, even for arbitrarily strong interactions. 

\subsection{ Model and diagrammatic expansion} 
As a starting point for the derivation of the new diagrammatic description, let us assume a Hamiltonian of a form that encapsulates the processes generally found in models of two-component lattice fermions and fermionized spin systems:
\bea
H_0=\hat{\mu},\;H_1=\hat{U}+\hat{J}+\hat{t}.\label{model}
\eea
Here, $\hat{\mu}$ is assumed to be local and bilinear, i.e., a chemical potential.
The term $\hat{U}$ is a contact interaction that is local and non-bilinear. 
The operator $\hat{J}$ describes a nonlocal interaction that is mediated by a Boson, like super-exchange or the nonlocal part of a Coulomb interaction term. Finally, $\hat{t}$ is nonlocal and fermionic, generally corresponding to hopping.

In principle, we can treat the model (\ref{model}) through expansion in $H_1$,
\bea
\langle\hat{o}\rangle\!= \!
Z^{\!-\! 1}\sum_n \!\frac{(\!-\! 1 \!)^n}{n!} \! \! \int_0^\beta \! \! d \tau_i  
\text{Tr}\{e^{\! \!-\beta \! H_0  }  T \![H_1(\!\tau_1 \!) ...  H_1(\!\tau_n \!)  \hat{o} ] \},\;\;\;\;\label{expansion}
\eea
and due to bi-linearity of $H_0$, the contractions can be evaluated using standard Matsubara formalism based on Wick's theorem \cite{fetter}. 
Furthermore, we note that the unperturbed theory is also local, so that all contractions of operators that are separated in space vanish a priori, i.e.
\bea
G^0_{\alpha\beta}(i-j,\tau)=G^0_{\alpha\beta}(\tau)\delta_{i,j}. \label{G0local}
\eea
At this stage, we aim to exploit the combination of bi-linearity and locality of the unperturbed theory. Thus, we first recall that Wick's theorem allows us to obtain an answer from the series of connected diagrams by cancellation of disconnected contributions and the partition function \cite{fetter}. 
Secondly, we notice, in accordance with (\ref{G0local}), that all calculations are carried out in the atomic limit, where the full expectation value of an operator is generally trivial to obtain, and does not even require the evaluation of diagrams. In particular, this allows the evaluation of certain classes of terms up to infinite order, thus paving the way for non-perturbative treatment of contact interactions, for example. 

When using these properties in conjunction, we do however face a fundamental problem in that the full contraction of a set of operators contains a mix of connected and disconnected topologies, which runs very much contrary to the concept of diagrammatic expansions. This complication is further bolstered because connectivity of a set of contractions is a nonlocal property. 
The principal solution to this problem is to divide the set of contractions on a lattice site $i$ into groups according to their connectivity, which effectively gives rise to a set of connected vertices that form the basis for an alternative diagrammatic technique.

\subsection{ Strong-coupling vertices}
Let us start by dividing the second part of the Hamiltonian (\ref{model}) into local and inter-site terms according to 
\bea
H_1=\hat{U}+H_I,\;\hat{U}=\sum_i \hat{U}_i,
\eea
where $i$ refers to lattice sites. 
Then, we proceed to introduce the following shorthand notation for the normalized time-ordered integration which appears in expansions of the form (\ref{expansion}):
\bea
\Gamma_n=\frac{(-1)^n}{n!}\int_0^\beta d\tau_1...d\tau_n T_\tau
\eea
with the generalization
\bea
\Gamma_n\Gamma_m=\frac{(-1)^{n+m}}{n!m!}\int_0^\beta d\tau_1...d\tau_{n+m} T_\tau.
\eea
We can now write the expansion in $H_1$ as 

\bea
\sum_n\Gamma_n H_1^n=\sum_{n,m} \Gamma_n\Gamma_m\hat{U}^nH_I^m\\
=\sum_{m,n_1,n_2...}\Gamma_m\Gamma_{n_1}... \hat{U}_1^{n_1}\hat{U}_2^{n_2}...H_I^m \label{spatialDecomp}
\eea
where the subscript of $U$ refers to lattice site and the string of operators $H_1^n$ are assumed to depend on $\tau_1...\tau_n$. 
Next, we introduce $\bar{O}_i$ to denote the set of operators on the site $i$ that are attributable to nonlocal terms in the Hamiltonian (i.e., $H_I$) or the measured operator $\hat{o}$. Expressing the expansion (\ref{expansion}) in this new language, we find
\bea
\sum_n \Gamma^n \sum_{\bar{x}} \Big\langle\!\prod_i \sum_{n_i} e^{-\beta H_{0,i}} \Gamma^{n_i}U_i^{n_i} \bar{O}_i\Big\rangle_c.\label{expansion2}
\eea 
Here, $n$ is the expansion order of the inter-site terms $H_I$, while $\bar{x}$ denotes their spatial degrees of freedom, which are summed over accordingly. The subscript $c$ implies connected topologies. 

Since the bare Greens function is local (\ref{G0local}), it follows that all contractions in (\ref{expansion2}) may be carried out locally also.  However, the problem that remains is that we are interested in topologies that are connected, and this very property is nonlocal. To overcome this difficulty, we cannot simply compute the local trace; instead, we have to sort all local contractions according to their connectivity. To do so, we begin by breaking out 
 the local terms on the site $i$ that are not connected to any of the operators $\bar{O}_i$:
\bea
\sum_{n} \langle \Gamma^{n} U_i^{n} \bar{O}_i\rangle_{\hat{\mu}}=\sum_{n,m} \langle \Gamma^n U_i^n \bar{O}_i\rangle_{\hat{\mu},e}\langle \Gamma^m U^m   \rangle_{\hat{\mu}}.\label{ext}
\eea
Here, the subscript $\langle \rangle_{\mu,e}$ denotes the subset of contractions such that all diagrammatic elements are connected to at least one external line (i.e., an operator that is attributable to a nonlocal process).
Discarding the disconnected topologies of (\ref{ext}) we may write (\ref{expansion2}) as
\bea
\sum_n \Gamma^n \sum_{\bar{x}}  \Big[\!\prod_i \sum_{n_i} \langle \Gamma^{n_i}U_i^{n_i} \bar{O}_i\rangle_{\mu,e}\Big]_c,\label{expansion3}
\eea 
where subscript $c$ once again implies the subset of these topologies that are connected. However, this is a nonlocal property that depends both on the contractions on a site and the inter-site processes that connect different sites. To make further progress, we have to sort the local contractions according to the connectivity of the set of operators $\bar{O}_i$. 

Thus, we define the {\it connected} set of contractions denoted by $\langle...\rangle_{\mu,c}  $ as the subset of $\langle...\rangle_{\mu,e} $ for which all elements of $\bar{O}_i$ are connected by local contractions. 
As an example, we may consider the case when $\bar{O}$ has only two elements:
\bea\nonumber
\sum_n\langle \Gamma^n U_i^n \hat{O}_1\hat{O}_2\rangle_{\hat{\mu},e}=
\sum_n\langle \Gamma^n U_i^n \hat{O}_1\hat{O}_2\rangle_{\hat{\mu},c}
\\
+\sum_{n,m}\langle \Gamma^n U_i^n \hat{O}_1\rangle_{\hat{\mu},c}\langle \Gamma^m U_i^m \hat{O}_2\rangle_{\hat{\mu},c}.\label{divide2}
\eea
Thus, we have sorted the contractions into two parts: Those where $O_1,O_2$ are connected, which defines the {\it Connected vertex}, and those where they are disconnected. 
In principle, we can generalize this procedure to the case of an arbitrary number of elements of $\bar{O}$ by constructing a recursion that is reminiscent of Determinant diagrammatic techniques \cite{PhysRevLett.119.045701,Rossi_2017}: 
For a given set of operators $\bar{O}$, we take as our starting point a set of local contractions of the form
\bea
\sum_n \langle \Gamma^n U^n \bar{O}\rangle_{\hat{\mu},e}. \label{localCont}
\eea
The set of connected topologies may be obtained by subtracting those that are disconnected. 
To list these, we begin by sorting them according to which of the operators $\bar{O}$ are connected to $O_1$, and denote this set by $A$. The set of 
contractions for which the operators in $A$ are connected to each other, but not to the remaining operators ($\hat{O}\setminus A$) is by definition given by
\bea
\xi_{\bar{O},A}
 \sum_{n,m} \langle \Gamma^n U^n A\rangle_{\hat{\mu},c}\langle \Gamma^m U^m \bar{O}\setminus A\rangle_{\hat{\mu},e},
\eea
where $\xi_{\bar{O},A}$ is a fermionic sign given by
\bea
\xi_{\bar{O},A}=(-1)^c
\eea 
where $c$ is the number of fermionic commutations associated with the reordering 
\bea
T_\tau \bar{O}\to T_\tau A \times T_\tau(\bar{O} \setminus A).
\eea
In the next stage, we have to sum over all possible choices of $A$. Here, we have two restrictions: Firstly, since $A$ is the set of operators connected to $O_1$ it follows that $A$ must contain $O_1$. Secondly, since we are only interested in the set of disconnected topologies, it follows that $\bar{O}\setminus A$ must be a nonempty set--otherwise, all operators are connected. Thus, $A$ is a {\it proper} subset of $\bar{O}$. Summing over all choices of $A$, we obtain
\bea
\sum_{A\subsetneq \bar{O}, \hat{O}_1\in A}
\xi_{\bar{O},A}
 \sum_{n,m} \langle \Gamma^n U^n A\rangle_{\hat{\mu},c}\langle \Gamma^m U^m \bar{O}\setminus A\rangle_{\hat{\mu},e}.\label{disconnected}
\eea
Using (\ref{disconnected}) we can then construct a recursive relation for the connected vertex on the form 
\bea\nonumber
V[\bar{O}]=\sum_n\langle \Gamma^n U_i^n \bar{O}\rangle_{\hat{\mu},c}=\sum_n\langle \Gamma^n U_i^n \bar{O}\rangle_{\hat{\mu},e}
\\
\!-\!\sum_{A\subsetneq \bar{O}, \hat{O}_1\in A}\!\xi_{\bar{O},A}
 \sum_{n,m} \langle \Gamma^n U^n A\rangle_{\hat{\mu},c}\langle \Gamma^m U^m \bar{O}\!\setminus\! A\rangle_{\hat{\mu},e},\label{recursion}
\eea
such that it can be directly computed from terms of the form (\ref{localCont}). We then recognise that (\ref{ext}) can be written
\bea
\sum_{n} \langle \Gamma^{n} U^{n} \bar{O}\rangle_{\hat{\mu}}=\sum_{n} \langle \Gamma^n U^n \bar{O}\rangle_{\hat{\mu},e} \Tr e^{-\beta (\hat{\mu}+\hat{U})}
\eea
implying 
\bea
\sum_n\langle \Gamma^n U_i^n \bar{O}\rangle_{\hat{\mu},e}=\langle \bar{O}\rangle_{\hat{\mu}+\hat{U}}\label{expO}. 
\eea
Here it should be pointed out that for the Hubbard model, the summation (\ref{expO}) has a finite convergence radius with respect to $U$ \cite{PhysRevLett.114.156402}. It is also known that this model possesses singularities in the irreducible vertex functions at strong interactions \cite{PhysRevB.98.235107}. 
However, the divergence of (\ref{expO}) can be removed by applying second fermionization \cite{PhysRevB.84.073102} to the Hubbard model. A proof of this is given in appendix A. 

Noting that the summation (\ref{expO}) may be rendered convergent and equated to an expectation value taken in the atomic limit, it can be computed exactly, and so the construction of the connected vertex (\ref{recursion}) is an exactly solvable problem. 
This allows the contact interactions to be treated completely non-perturbatively.

We also note that any set of the form (\ref{localCont}) may be decomposed into a sum of sets of connected vertices:
\bea
\sum_n\! \langle \Gamma^n U^n \bar{O}\rangle_{\hat{\mu},e}\!=\!V[\bar{O}]\!+\!\!\sum_{\bar{O}'}\!\xi_{\bar{O},\!\bar{O}'}\! V[\bar{O}']V[\bar{O}\!\setminus\! \bar{O}']...
\eea
Inserting this into (\ref{expansion3}) we find that all topologies in the expansion may be expressed in terms of connected vertices. Likewise, any set of connected vertices may be expressed in terms of contractions of the form (\ref{expansion3}). Therefore, we can expand directly in connected topologies of connected vertices:

\bea
\sum_{N}\! \Big[\!\sum_{\alpha_1}\! V_{\alpha_1}(x_1,\bar{\tau_1})\!...\!\sum_{\alpha_N}\! V_{\alpha_N}(x_N,\bar{\tau_N}) \eta(\alpha_1\!...\alpha_N)\!\Big]_c\!,\label{Vseries}
\eea
where $\eta$ is the normalisation. In the next section, we will establish rules for the expansion (\ref{Vseries}). 

\subsection{ Diagrammatic rules}
To obtain connected topologies, the connected vertices defined by the recursion (\ref{recursion}) have to be connected via external lines that originate in the nonlocal operators, i.e. $\hat{t},\; \hat{J}$. 
However, a complication that remains is determining the sign of a contribution, which generally requires the construction of diagrammatic rules that govern the expansion. 
Since this derivation is essentially based on Wick's theorem, we first have to make contact with Feynman type diagrammatics in order to derive the corresponding principles for the strong-coupling expansion. 

\begin{figure}[!htb]
\includegraphics[width=\linewidth]{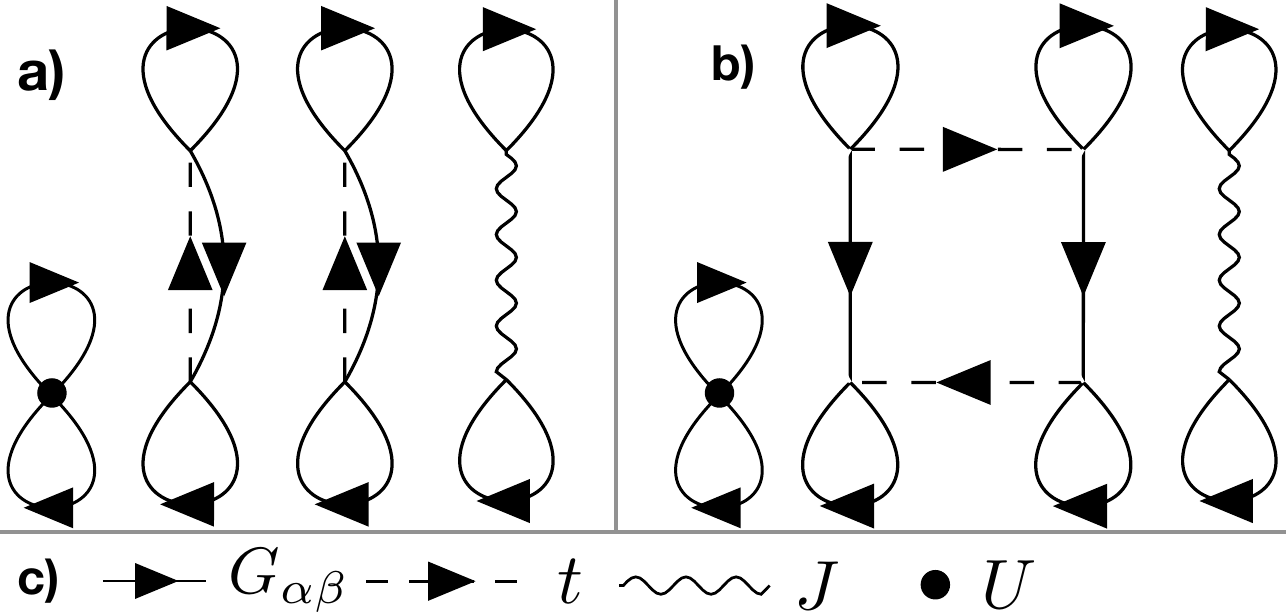}
\caption{
{\bf Reference contractions.}
Given a set of operators, we can define a {\it reference contraction} (a) where, all operators are contracted with its natural partner. While generally not a connected topology, the fermionic sign of the reference is positive. Swapping a set of operators being contracted gives rise to a fermionic sign, and so the diagram (b) possesses a negative prefactor. 
This could in principle also be achieved by changing the connectivity of the $t-$lines, which is thus an equivalent operation. 
}
\label{refcont}
\end{figure}

In standard literature, the expansion is typically conducted in conventional two-body interactions, and the overall sign of a diagram is generally expressed in terms of the number of fermionic loops \cite{fetter}. 
Proceeding to more general models that for example include projected hopping, the resulting Feynman rules must generally be obtained from Wick's theorem. 
A convenient way of doing this is to introduce a {\it reference contraction}:
Specifically, we understand that we can write a fermionic theory on a form where creation and annihilation operators form natural pairs, whose contraction corresponds to an infinitesimally backward propagating fermion. Thus, for an expression of the form
\bea
U  n_{i\uparrow} n_{i\downarrow}\; J  n_{i\sigma} n_{i\sigma'}  \;t n_{j\uparrow}  c^\dagger_{j\downarrow} c_{k\downarrow} n_{k\uparrow}...\;\;
n_{i\sigma}=c^\dagger_{i\sigma} c_{i\sigma},\label{operatorstring}
\eea
every operator is contracted with its natural partner to form a diagrammatic element as shown in Fig. (\ref{refcont}, a), for which the fermionic sign is positive. Swapping the operators being contracted (Fig \ref{refcont}, b) gives rise to a fermionic sign, and so all diagram topologies can be characterized by whether they are related to the reference by an even or an odd number of such swaps. 

\begin{figure}[!htb]
\includegraphics[width=\linewidth]{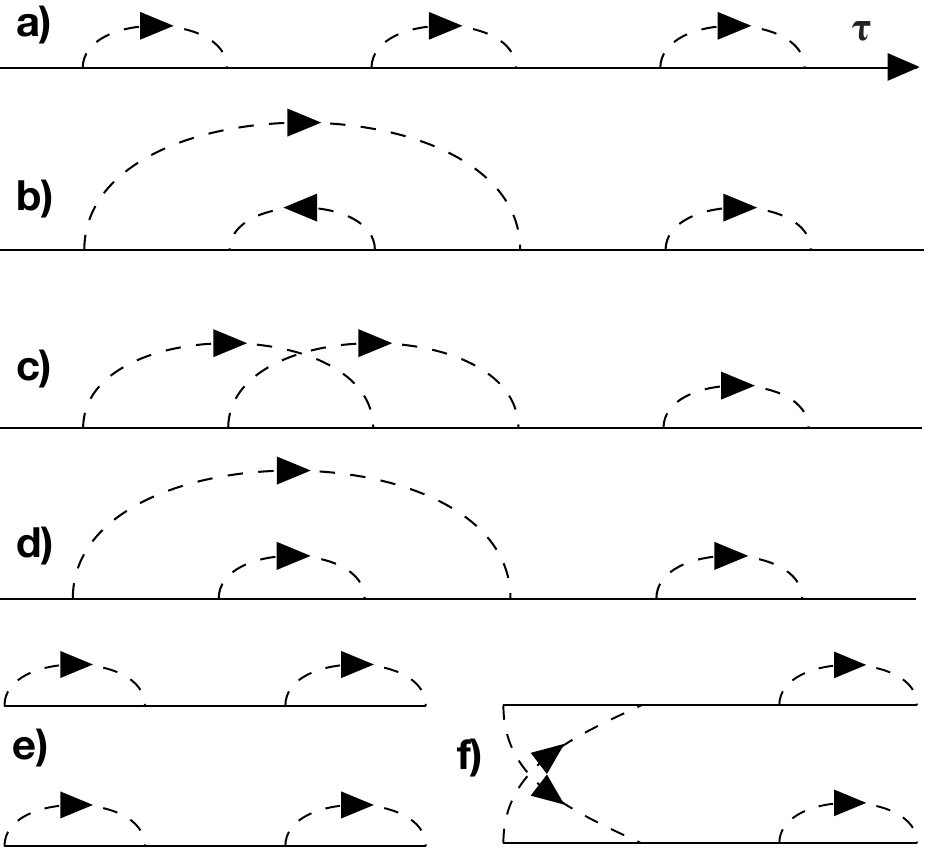}
\caption{
{\bf Diagrammatic rules.}
The reference contraction of a connected vertex (a) is obtained by taking the external lines to be non-intersecting, have no time-overlap, and be forward-propagating in the case of fermions. In our notation, the horizontal line corresponds to imaginary time, and so forward propagation implies that the external line is traveling from left to right.
 To generate arbitrary diagram topologies from reference vertices, we require two basic updates: {\it Swapping} the connectivity of two external lines, we can go from (a) to (b), whilst {\it commuting} the operator order takes us from (a) to (c).  
The diagrams (c) and (d) are related by alternatively a swap {\it or} a commute operation. 
Swapping external lines also allows us to connect different vertices, such as going from (e) to (f).
}
\label{scd}
\end{figure}

Adapting this idea to the strong-coupling expansion, the first natural stage is to define a reference contraction for the connected vertex. While there are several equivalent ways of doing this, the simplest choice is arguably to consider a vertex where all external lines are temporally non-overlapping, non-intersecting, and in the case of fermionic lines, also forward propagating in time, see Fig. (\ref{scd}, a). 
To confirm that this diagram indeed carries a positive fermionic sign, we simply note that from the underlying operators, we can form the Feynman reference contractions of the type (\ref{refcont}, a) without commuting any of them. Thus, if we, for example, assume that the external lines in Fig. (\ref{scd}, a) are fermionic, then we obtain an operator product of the form 
\bea
\sim c_\alpha^\dagger c_\alpha   c_\beta^\dagger c_\beta c_\gamma^\dagger c_\gamma.\label{fermiRef}
\eea
Summing up all contractions of (\ref{fermiRef}) in accordance with the underlying Feynman series (including disconnected topologies), this is equivalent to the expectation value of the operators, corresponding to a positive fermionic sign.

As illustrated in Fig. \ref{scd}, we require two basic updates to generate arbitrary diagrams from a set of reference contractions, namely {\it swapping} the connectivity of two external lines, and {\it commuting} operators within a vertex. 
For fermionic lines or operators, particle statistics suggest that these operations are odd, and this is indeed what transpires from the underlying Feynman type diagrammatics: 
Firstly, swapping the connectivity of two fermionic lines is equivalent to changing the connectivity of an odd number of fermionic propagators, which according to Wick's theorem, is an odd operation. As an example, we may consider the operation (a $\to$ b) in Fig. \ref{refcont}, which could alternatively be realized by a swap of the fermionic operators {\it or} the $t-$lines. 
Secondly, the process (c$\to$ d) in Fig. \ref{scd} can be achieved with either a swap or a commute, implying that these operations have the same parity. 
By the same logic, operations on bosonic external lines do not give rise to a sign.

\subsection{ Analytic structure of the connected vertices}
Whilst the recursion (\ref{recursion}) provides a principal definition of the diagrammatic elements of the expansion, computing, and also storing these objects in memory is only possible given an efficient representation. 
In particular, for a vertex with $N$ external lines, the naive description yields $N-1$ imaginary-time differences and equally many dimensions of the mathematical object to be constructed and stored, so that the task quickly becomes intractable.

The solution to this problem can be found by noting that in the recursion (\ref{recursion}), the vertex is expressed in expectation values of the form (\ref{expO}), which are taken with respect to the entire local part of the Hamiltonian, i.e. $H_L=\hat{\mu}+\hat{U}$. 
If we express the nonlocal part $H_I$ in an operator basis which possesses a trivial time-evolution with respect to $H_L$, then the time-dependence of the entire vertex becomes equally simple.
The specific representation which allows this to be achieved, and which notably also forms the starting point for derivation of the t-J model \cite{PhysRevB.18.3453}, takes the form
\bea\nonumber
c_{i\sigma}^\dagger=d_{i\sigma}^\dagger + h_{i\bar{\sigma}},\;d_{i\sigma}^\dagger=c_{i\sigma}^\dagger n_{\bar{\sigma}},\;h_{i\bar{\sigma}}=c_{i\sigma}^\dagger(1-n_{i\bar{\sigma}}),\;\;\\
c_{i\sigma}=d_{i\sigma} + h^\dagger_{i\bar{\sigma}},\;d_{i\sigma}=c_{i\sigma} n_{\bar{\sigma}},\;h^\dagger_{i\bar{\sigma}}=c_{i\sigma}(1-n_{i\bar{\sigma}}),\;\;
\label{newOperators}
\eea
where $\bar{\sigma}=-\sigma$, while the corresponding time-dependence with respect to $H_L$ is given by
\bea\nonumber
d_{i\sigma}^\dagger(\tau)=e^{\tau H_L} d_{i\sigma}^\dagger e^{-\tau H_L}=e^{(U-\mu)\tau}d_{i\sigma}^\dagger\\\nonumber
h_{i\sigma}(\tau)=e^{\tau H_L} h_{i\sigma} e^{-\tau H_L}=e^{-\mu\tau}h_{i\sigma} \\\nonumber
d_{i\sigma}(\tau)=e^{\tau H_L} d_{i\sigma} e^{-\tau H_L}=e^{-(U-\mu)\tau}d_{i\sigma}\\
h_{i\sigma}^\dagger(\tau)=e^{\tau H_L} h_{i\sigma}^\dagger e^{-\tau H_L}=e^{\mu\tau}h^\dagger_{i\sigma}.
\label{CTIME}
\eea
Given a set of operators of the form (\ref{newOperators}) that are evaluated with respect to $H_L$, we can use (\ref{CTIME}) to divide it into a scalar part, which consists of an analytic function and an operator part which only depends on the order of the terms, according to
\bea
O_1(\tau_1)...O_N(\tau_N)
=f(\{\tau_i\}) 
 O_1...O_N.\label{vertexstructure}
\eea
Since the recursion (\ref{recursion}) implies that the connected vertex can be expressed in terms of expectation values of the form (\ref{expO}), it follows that we can break out the scalar part from this expression, and thus obtain an object of the form
\bea
V[\bar{O}(\{\tau_i\})]=f(\{\tau_i\}) V[O_1O_2...O_N]\label{analyticStructure}
\eea
where $f$ is an analytic function, while $V[O_1O_2...O_N]$ is a constant which only depends on the order of the operators, and correspondingly may be stored as a single floating point. 
Furthermore, we may note that using the basis (\ref{newOperators}) and exploiting the property (\ref{analyticStructure}), (\ref{expO}) essentially corresponds to the expectation value of a projection operator, which can be calculated exactly, and so the connected vertex is naturally obtained to machine precision. 

Finally, let us comment on the prelusive question about the feasibility of storing the vertices in lookup tables: For the Hubbard model, the basis  (\ref{newOperators}) gives a total of $8$ operators, implying that the number of vertices scales as $8^N$ where $N$ is the number of external lines or legs.
At $N=10$, this gives $\sim 10^9$ vertices, which translates to approximately $8$ GB at double precision. 
For the Heisenberg model, which can be described by only $4$ operators, we can afford to store all vertices up to $N=15$ with the same resources. Generally, Gutzwiller-projected systems will perform better than the Hubbard model in this respect. 
Exploiting symmetries and the fact that most vertices actually vanish due to particle and spin conservation, it might be possible to store somewhat larger objects.

 \begin{figure}[!htb]
\includegraphics[width=\linewidth]{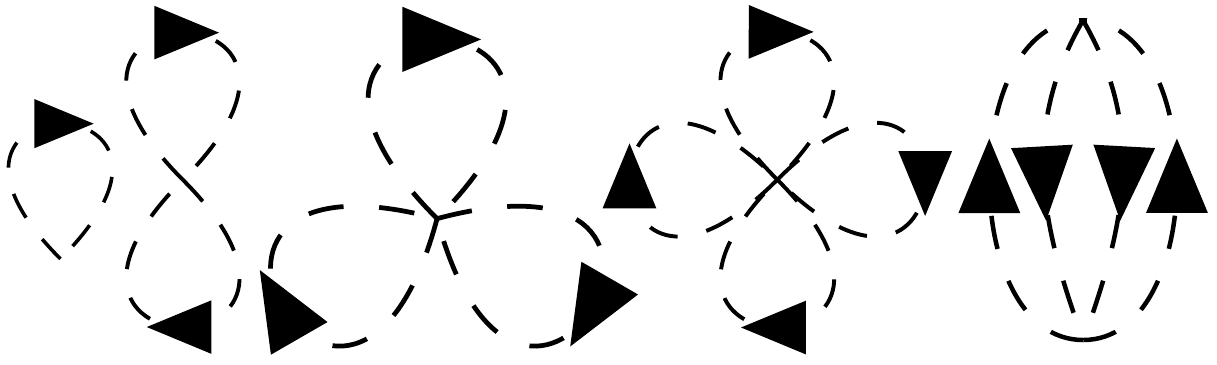}
\caption{
{\bf Strong-coupling diagrams.}
The set of topologies obtained for the Hubbard model up to order $N_t=4$ when boldifying the $t-$lines. Each dashed line represents a dressed hopping integral. At an expansion order $N_t$, the largest vertex that can be constructed has $2N_t$ external lines or legs. 
Using fermionization techniques and conventional diagrammatics, the number of topologies at the same expansion order can be estimated to $\sim 10^6$. 
}
\label{hubbardTop}
\end{figure}

\subsection{Observables} 
\begin{figure}[!htb]
\includegraphics[width=\linewidth]{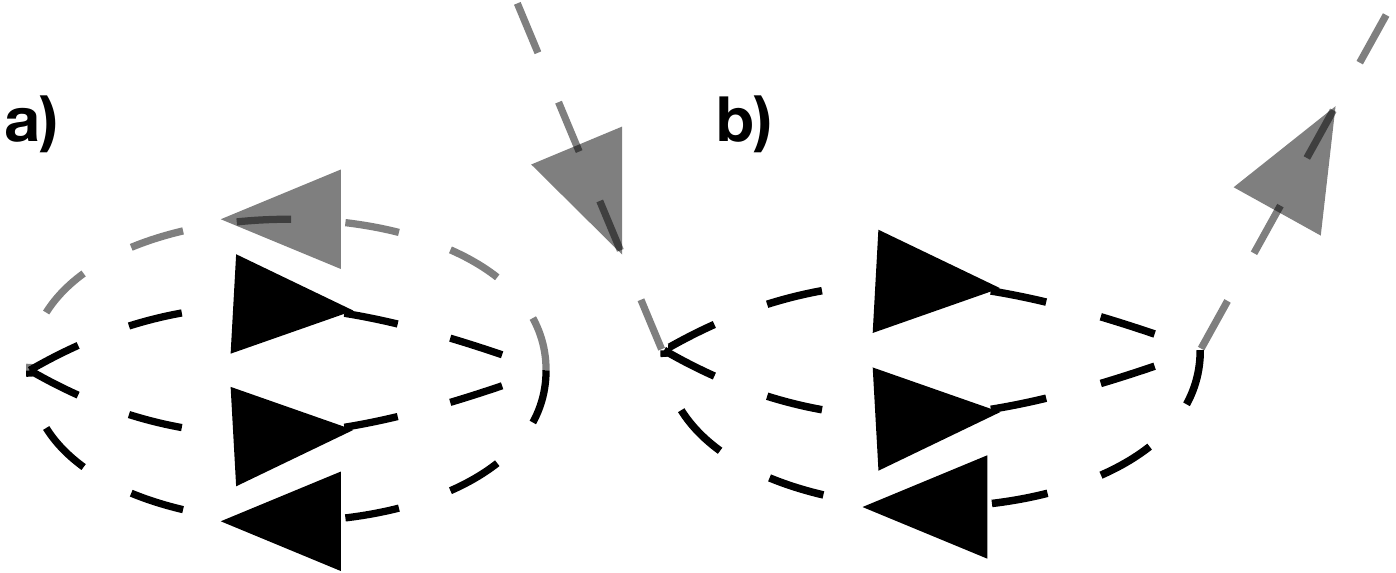}
\caption{
{\bf Extracting observables.} The shaded line in {\bf(a)} is tagged as a measuring line.
The resulting topology is then interpreted as if this was an external line {\bf(b)}, and the remaining part of the diagram gives a contribution to the polarization operator of the line type in question. 
}
\label{mline}
\end{figure}

In diagrammatic Monte Carlo, the extraction of observables is typically achieved using a measuring line as illustrated in Fig. \ref{mline}, see also \cite{Van_Houcke_2010}. One of the lines is then tagged and treated as an entrance and exit of a particle from the system, while the remains of the diagram are interpreted as a contribution to the self energy or the polarization, depending on the line type being considered. 
In the strong-coupling expansion, the particle propagators are hidden inside the connected vertices, and we only have access to the external lines that originate in the nonlocal processes. Therefore, the Greens function must be obtained from the polarization of the $t$-line, as opposed to via Dysons equation:
\bea\nonumber
G(\omega,\bk)=\Pi(\omega,\bk)+\Pi(\omega,\bk)t(\bk)\Pi(\omega,\bk)+...
\\
\implies G(\omega,\bk)=\frac{1}{\Pi^{-1}(\omega,\bk)-t(\bk)},\label{Gdressed}
\eea
where $\Pi$ is the polarization operator of the $t-$line. In spin models, two-point correlations can be computed from the polarization of the $J$-line, while access to further observables that do not correspond to any specific external line can in principle, be obtained by constructing appropriate operators solely for the purpose of measuring. Using multiple measuring lines, many-point correlators can be obtained.

\subsection{Benchmarks for the Hubbard model}
\begin{figure*}[!htb]
 \hbox to \linewidth{ \hss
\includegraphics[width=\linewidth]{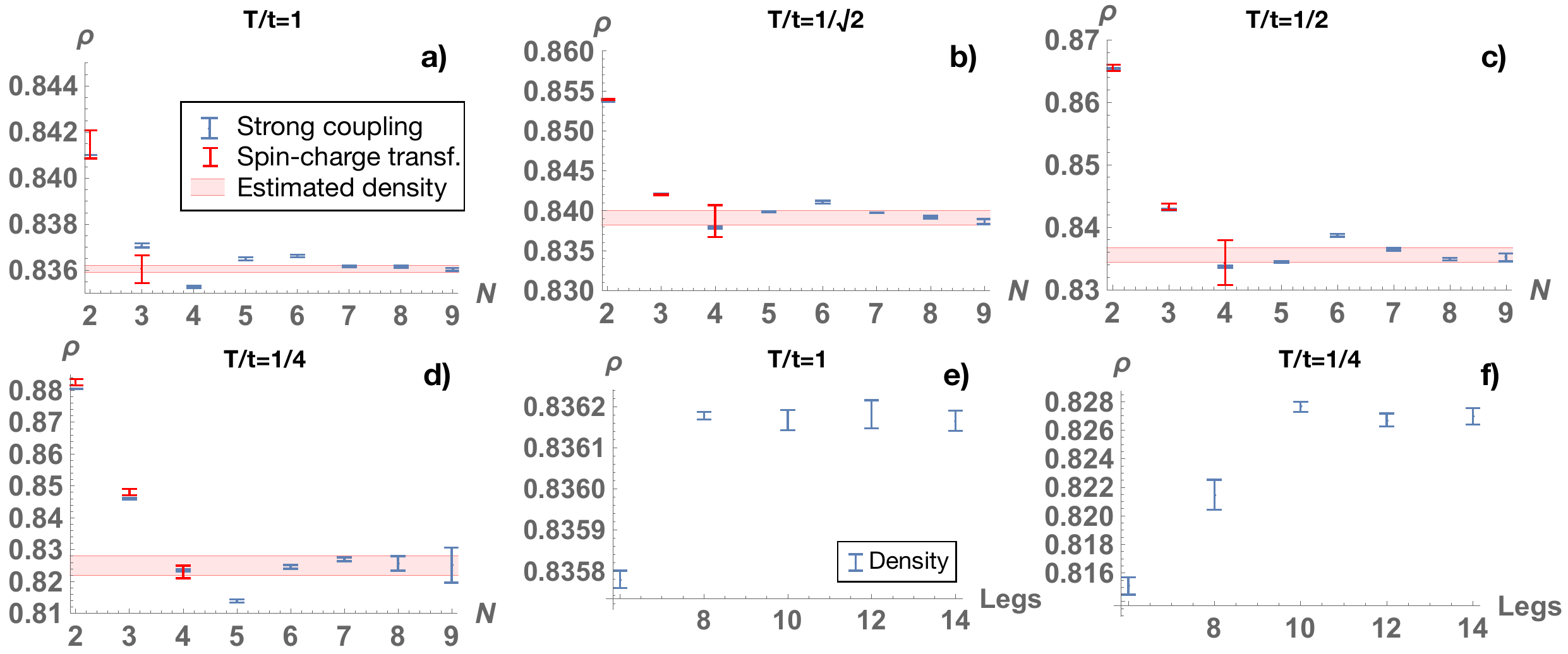}
 \hss}
\caption{
{\bf Series expansion for the carrier density.} In {\bf (a-d)} the filling factor is given as a function of expansion order, in the temperature range $1/4\le T/t\le 1$ for $\mu/t=2$ and $U/t=\infty$. The blue bars correspond to strong-coupling theory (this work), while the red bars were obtained from conventional diagrammatic treatment of the spin-charge transformed Hubbard model \cite{PhysRevB.97.075119}. Since both these methods rely on a bold expansion in $t$ we expect them to provide identical results, and this holds true within or almost within error bars. 
The shaded region gives an estimate of the density at $N=\infty$, though a precise extrapolation to infinity is not possible at this stage. 
The vertex size is truncated at $16$ legs, which affects the last term when $N=9$. 
In {\bf (e-f)} we examine the effect of truncating the vertex size at $T/t=1$ and $T/t=1/4$ respectively. We thus set the expansion order in $t$ to $N=7$, and observe how the predicted carrier density varies with the maximal number of vertex legs. At $T/t=1$, the corrections beyond $8$ legs vanish within the error bars, which are of the order $\sim 3\times 10^{-5}$. At $T/t=1/4$, the corrections beyond $10$ legs falls within the error bar of $\sim 10^{-3}$. Hence, at $N=9$, the error due to truncation of the vertex size can be expected to be vanishingly small compared to statistical noise.   
}
\label{results}
\end{figure*}

To evaluate the strong-coupling method outlined above, we will now compare it to results obtained with two other state of the art numerical protocols. 
For the Hubbard model, NLCE can produce unbiased results in the strong-coupling regime, including $U=\infty$ \cite{PhysRevE.89.063301}. This technique is exact in the limit of infinite cluster size, and correspondingly it is also controllable, as convergence of the result with respect to cluster size can be readily checked. 
A second method that is also applicable in this case is SCT-DMC \cite{PhysRevB.97.075119}, which is based on spin-charge transformation and a skeleton expansion in the hopping integral $t$. This also has an additional benefit: Since strong-coupling theory and SCT-DMC share the same expansion parameter and rely on identical skeleton schemes, they are comparable on an order by order basis. 

When calculating the strong-coupling expansion, there are two principal computational efforts: 
Firstly, the connected vertices have to be obtained from the recursion (\ref{recursion}). In practice, this set has to be truncated at some given vertex size. We were able to obtain all vertices possessing up to $16$ operators attributable to nonlocal processes, which translates to as many external lines, or legs. Since the nonlocal terms $\sim t_{ij} c^\dagger_i c_j$ each posses two operators, the largest vertex that can be constructed from $N_t$ nonlocal terms has $2N_t$ legs, see also Fig. (\ref{hubbardTop}). 
Secondly, the observables have to be extracted from an expansion in $t$ using the connected vertices. Here, truncation of the total number of nonlocal terms $N_t$ is also necessary. We used a standard worm protocol \cite{Van_Houcke_2010} that is very similar to that of \cite{PhysRevB.97.075119}, and were able to reach an order of up to $N_t=9$.  
At this order it is in principle possible to create a connected vertex with as many as $18$ legs, and the fact that we had to truncate the vertex size thus affects the last term. 
Summation was conducted for a chemical potential of $\mu/t=2$ and an infinite onsite repulsion $U=\infty$ in a temperature range $1/4\le T/t\le 1$.

A summary of the result is given in Fig. (\ref{results}). Comparing these to SCT-DMC data, we obtain a first confirmation that strong-coupling treatment indeed reproduces results from Feynman type diagrammatics. The two series are in good agreement in all cases where results can be obtained to sufficiently high accuracy. 
What also transpires from this comparison is the disparity in efficiency of the two methods. 
We are now able to reach $N_t=9$, while the SCT-DMC results are limited $N_t=4$ or less. This improvement is very significant considering that the computational complexity scales factorially with expansion order. 

To examine the impact of truncating the vertex size, we also compute the equation of state at an order of $N_t=7$, and vary the maximal vertex size in the span $6\le N_{\text{legs}}\le 14$. At $T/t=1$, we estimate that truncation at $8$ legs gives an error of less than $3\times 10^{-5}$, though no lower bound could be set, as this is within error bars. At $T/t=1/4$, we estimate that truncation at $10$ legs gives an error of less than $10^{-3}$, with no lower bar. This implies that the truncation error present at $N_t=9$ in the results presented here is small compared to the statistical noise. It does also indicate that the current results are limited by the expansion order rather than achievable vertex size. 

\begin{figure}[!htb]
\includegraphics[width=\linewidth]{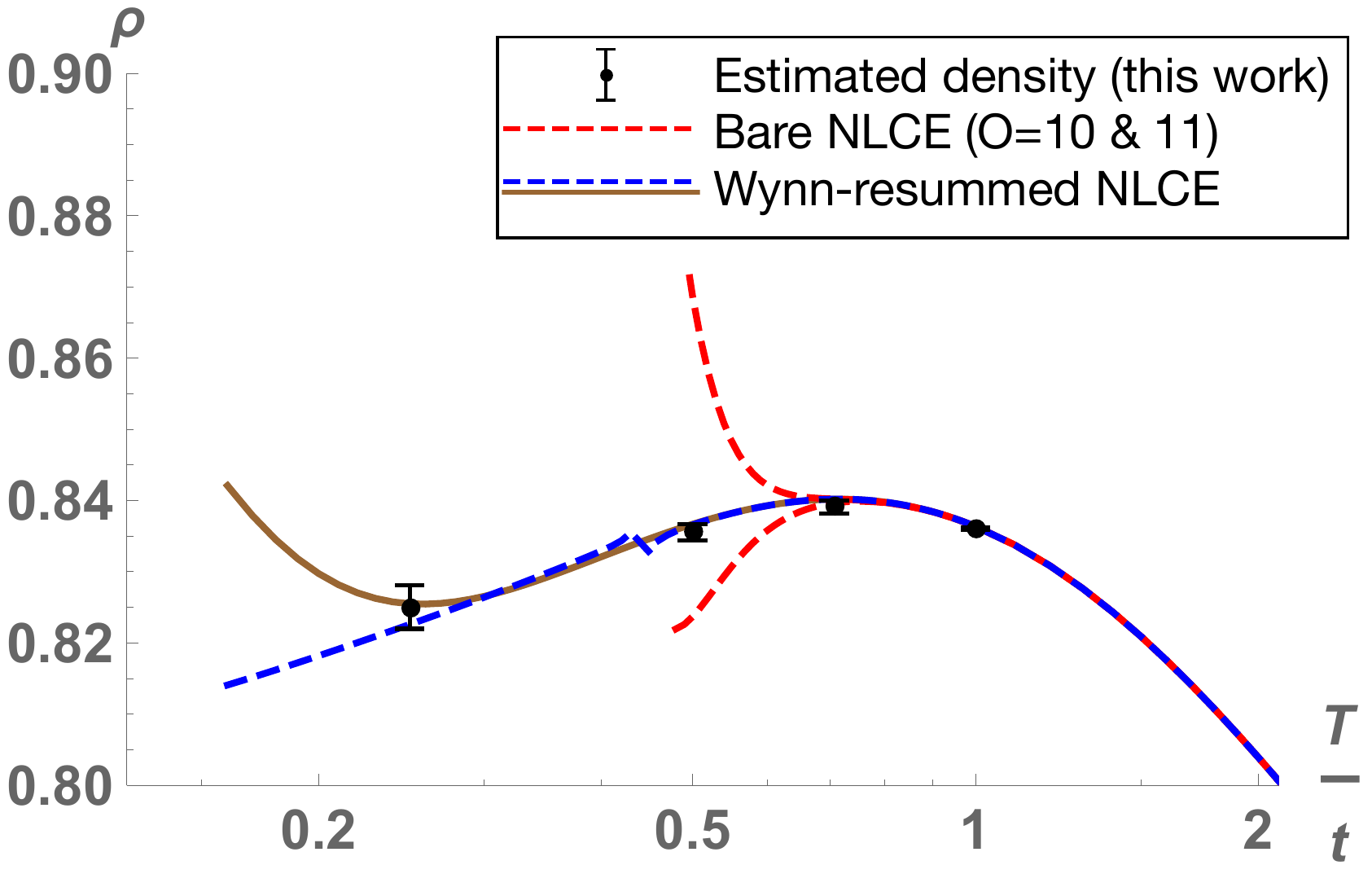}
\caption{
{\bf Comparison to NLCE.} The black bars give estimates of the filling factor obtained from strong-coupling treatment for $\mu/t=2$, $U/t=\infty$, see also Fig. \ref{results}. The red dashed lines show NLCE data at orders $10$ and $11$ respectively, which converge down to $T/t\approx 1/\sqrt{2}$ \cite{PhysRevE.89.063301}. Beyond this point, comparison has to be made with Wynn resummed NLCE (dashed blue and solid brown lines). The results are in good agreement. 
}
\label{compare}
\end{figure}

Extracting estimates for the equation of state from strong-coupling theory, we can compare these to NLCE, see Fig. (\ref{compare}). For temperatures $T/t\ge 1/2$, we observe excellent agreement that is within moderate error bars. At the lowest temperature, the uncertainty increases and the NLCE data also begins to diverge, but the results still agree within error bars.

Further improvement to this method can be made by tailoring new sampling protocols specifically to this expansion. Very significant gains can also be made by altering the structure of the series itself. Shifted action \cite{PhysRevB.93.161102} or generalizations thereof \cite{kim2020homotopic} can be employed to improve the rate of convergence of the series, both with respect to overall expansion order, as well as vertex size.

\subsection {Summary and outlook}
In conclusion, we have derived a diagrammatic technique based on connected strong-coupling vertices, that can be applied to lattice fermions and quantum spin models.  
This method allows large, or even infinite, contact interactions to be treated non-perturbatively, thus overcoming a longstanding obstacle for diagrammatic methods in the strongly correlated regime. 
For the Hubbard model, we are able to obtain self consistent solutions up to an expansion order of $N_t=9$, that display good agreement with results from NLCE. 

Experimental progress with strongly correlated systems--using ultracold atomic gases--is now rapid.
With quantum gas microscopy \cite{Gross995,Chiu251}, key features of the doped Mott insulator can be observed at the single particle level, and at temperatures where spin-correlations become significant \cite{Mazurenko2017,Koepsell2019,koepsell2020microscopic,Cheuk1260,Parsons1253}. 
Strong coupling Diagrammatic Monte Carlo can be used to extract virtually exact correlators at low temperatures, that can be directly compared to these experiments.

\subsection{ Acknowledgments}
This work was supported by the the Swedish Research Council (VR) through grant 2018-03882. Computations were performed on resources provided by the Swedish National Infrastructure for Computing (SNIC) at the National Supercomputer Centre in Linköping, Sweden.
The author would like to thank Marcos Rigol for providing NLCE results for benchmarking as well as Kun Chen, Boris Svistunov and Nikolay Prokof'ev for important input and discussions.

\bibliography{biblio.bib}

\section{Appendix A: Summation over contact interactions}
The summation over all contractions on the site $i$ such that all diagrammatic elements are connected to at least one external line, is given by Eq. (\ref{expO}), i.e.
\bea
\sum_n\langle \Gamma^n U_i^n \bar{O}\rangle_{\hat{\mu},e}=\langle \bar{O}\rangle_{\hat{\mu}+\hat{U}}\label{expO_APP}. 
\eea
We begin by noting that the set of operators $\bar{O}$ may be expressed in the operator basis (\ref{newOperators}) as follows:
\bea
\bar{O}=\sum_\alpha \bar{O}_\alpha,\; \langle \bar{O}\rangle_{\hat{\mu}+\hat{U}}= \sum_\alpha \langle \bar{O}_\alpha\rangle_{\hat{\mu}+\hat{U}} \label{stringDecomp}
\eea
where $\bar{O}_\alpha $ is a set of operators of the form (\ref{newOperators}). For a finite set  $\bar{O}$, we furthermore note that the range of $\alpha$ is also finite. 
Using (\ref{CTIME}) we obtain
\bea\nonumber
\langle \bar{O}_\alpha\rangle_{\hat{\mu}+\hat{U}}= \Tr e^{-(\hat{U}+\hat{\mu})} O_{\alpha,1}(\tau_1)...O_{\alpha,N}(\tau_N)\\
= f(\{\tau_i\})  \Tr e^{-(\hat{U}+\hat{\mu})} O_{\alpha,1}...O_{\alpha,N}, \label{projection}
\eea
where $f(\{\tau_i\})$ gives the time dependence in accordance with (\ref{analyticStructure}), and $H$ is expressed in units of temperature. 
For the Hubbard model, (\ref{projection}) is analytic on the real axis, but not in the entire complex plane due to zeros of the partition function that occur for complex values of $U$, and so the convergence radius is finite when expanding in contact interactions. 

To solve this problem, we use second fermionization to construct a dual representation which is free of a large expansion parameter, and thus possesses a convergent series regardless of model parameters. A detailed discussion of fermionization techniques is given in \cite{PhysRevB.84.073102}, but we will here recapitulate the central ideas of this approach:
Essentially, the goal is to remove the doublons from the trace, and then reintroduce them as hard core bosons that are subsequently fermionized. The end results of this procedure is that the contact interaction becomes a bilinear term in the Hamiltonian.

First, we thus remove the doublons entirely by introducing a projection operator $p_G$ and an auxiliary fermionic field with the number operator $n_A$:
\bea
H=-\mu (n^e_\uparrow+n^e_\downarrow)+ p_G,\;\;\; 
p_G=n^e_\uparrow n^e_\downarrow i\pi n_A,\label{gutzwiller}
\eea
where $n^e_\sigma$ are electron number operators. When we trace over $n_A=0,1$, the configurations for which $n^e_\uparrow n^e_\downarrow=1$ obtain an imaginary energy shift of $0 $ or $i\pi$ respectively which in turn give them opposite sign in the trace, such that the contribution vanishes. 

We then proceed to reintroduce the doublon in the form of a hard core boson, with an energy $U-2\mu$. 
The boson can in turn be fermionized, and thus gives rise to two fermionic components with number operators given by $n^d_0, \;n^d_1$. The state space correspondence is given by
\bea\nonumber
|n_{\text{boson}=0}\rangle\to |n^d_0=1,n^d_1=0\rangle,\\
|n_{\text{boson}=1}\rangle\to |n^d_0=0,n^d_1=1\rangle.\label{fermiboson}
\eea
The remaining states in the construction (\ref{fermiboson}) which correspond to $n^d_\uparrow+n^d_\downarrow\not=1$ has no physical counterpart, and are thus removed from the trace by the introduction of a Popov-Fedotov projection term \cite{0953-8984-29-38-385602} of the form 
\bea
p_D=(n^d_\uparrow+n^d_\downarrow-1)\frac{i\pi}{2},
\eea
such that the contribution from $n^d=0,\;2$, obtain a complex phase in the  in the trace and cancel. Finally, we are required to project out configurations where $n^e_\uparrow+n^e_\downarrow=1,\;n^d_1=1$, as this has no correspondence in the original state space. This can be achieved by 
\bea
p_H= (n^e_\uparrow-n^e_\downarrow)\Big(\frac{n^d_\uparrow-n^d_\downarrow}{2}+\frac{1}{2}\Big) i\pi n_A.
\eea
Including also the energy scale of the doublon, we thus arrive at a dual description of the local Hamiltonian according to
\bea
H\!=\!-\!\mu n^e\!+\!\Big(\frac{n^d_\uparrow\!-\!n^d_\downarrow}{2}\!+\!\frac{1}{2}\Big)E_D\!+\!p_G\!+\!p_D\!+\!p_H,\label{secondHubbard}
\eea
where $E_D=U-2\mu$ is the doublon energy. The partition function of (\ref{secondHubbard}) is given by
\bea
Z=2+2e^{-E_D}+4e^{\mu}\label{Z2}
\eea
which is indeed the partition function of the Hubbard model in the atomic limit, except for a trivial factor $2$ which we obtain when tracing over  the auxiliary field. 
In (\ref{secondHubbard}), the contract interaction is described by a bilinear term, and expansion is instead conducted in the projection operators $p_G,\; p_H$.

To examine the analyticity of the density matrix as a function of the expansion parameter, we parameterize the expansion terms $p_G,\;p_H\to \xi p_G,\;\xi p_H$ such that $\xi=1$ corresponds to the fully projected system. For convergence of the series, we then require analyticity within the unit circle $|\xi|\le 1$, regardless of model parameters. Next, we recall that the density matrix takes the form
\bea
\rho=\frac{W_i}{Z},\; W_i=e^{-E_i},\; Z=\sum_i e^{E_i}.
\eea
For finite model parameters, $W_i$ and $Z$ are analytic, implying that the density matrix is also analytic for non-vanishing $Z$. Correspondingly, demonstrating convergence of the series translates to ruling out zeros of $Z(\xi)$ within the unit circle $|\xi|\le 1$, which we will now do:

We begin by expressing the partition function in terms of $\xi$
\bea
Z(\xi)=a+b e^{-i\pi \xi}+ c e^{i\pi \xi}\label{partition}
\eea
with
\bea
a=2 e^{\mu -E_D}\!+\!e^{2 \mu -E_D}\!+\!2 e^{-E_D}\!+\!4 e^{\mu }\!+\!e^{2 \mu }\!+\!2,\\
   b=e^{\mu -E_D}+e^{2 \mu -E_D}+e^{2 \mu },\;
   c=e^{\mu -E_D},
\eea 
where in particular we note that
\bea
a> b+c. \label{scaleofabc}
\eea
Then we observe that
\bea
 Z(\xi\in \mathbb{I})>0, \label{imag-axis}
\eea
since the exponents in (\ref{partition}) are real on the imaginary axis. Furthermore we note that on the real axis, the exponentials in (\ref{partition}) only provide a phase, which together with (\ref{scaleofabc}) implies
\bea
 | Z(\xi\in \mathbb{R})|>0 \label{real-axis}
\eea
and so there are no poles on the real axis either. 

Away from the axes, the partition function is generally complex. 
For the imaginary part to vanish, we require 
\bea
b e^{\pi \xi_I }\sin (\pi\xi_R)=c e^{-\pi \xi_I }\sin (\pi\xi_R),\;\xi=\xi_R+i\xi_I.\label{imZ}
\eea
This equation has two types of solutions: Firstly, we have $\xi_R=0,\;\xi_R=\pm 1$, but these lie on the axes since $|\xi|\le 1$.
Secondly, we have a solution corresponding to
\bea
b e^{\pi \xi_I }=c e^{-\pi \xi_I }\implies e^{\pi \xi_I}=\frac{1}{\sqrt{1+e^\mu + e^{E_d+\mu}}}.\;\;\label{solution}\label{piXi}
\eea
Inserting (\ref{piXi}) into (\ref{partition}) we obtain
\bea
Z=a+2 e^{\mu-E_D}\sqrt{1+e^\mu+ e^{E_D+\mu}}\cos \pi\xi_R>0
\eea
for all real parameter values. 
Thus, we conclude that the density matrix is analytic within the unit circle $|\xi|\le 1$, and that (\ref{secondHubbard}) is described by a convergent series. 
Expressing the operators (\ref{newOperators}) in the basis $n_\downarrow,n_\uparrow,d_0,d_1$ we obtain a convergent summation in Eq. (\ref{expO}).

\end{document}